\documentclass[10pt,conference]{IEEEtran}
\usepackage{cite}
\usepackage{amsmath,amssymb,amsfonts}
\usepackage{algorithmic}
\usepackage{graphicx}
\usepackage{textcomp}
\usepackage{xcolor}
\usepackage[hyphens]{url}
\usepackage{fancyhdr}
\usepackage{hyperref}

\usepackage{needspace}
\usepackage{enumitem}
\usepackage{listings}
\usepackage{tcolorbox}
\usepackage{ifthen}
\usepackage{subcaption}
\usepackage{siunitx}
\usepackage{makecell}
\usepackage{placeins}

\setlist[itemize]{leftmargin=11pt}

\newcommand{\revision}[1]{\textcolor{black}{#1}}
\newcommand{\shepherd}[1]{\textcolor{black}{#1}}

\title{Nugget: Portable Program Snippets}

\newcommand\hpcaauthors{Zhantong Qiu, Mahyar Samani, Jason Lowe-Power}
\newcommand\hpcaaffiliation{University of California Davis}
\newcommand\hpcaemail{\{ztqiu, msamani, jlowepower\}@ucdavis.edu}

\def\aeopen{}           
\def\aereviewed{}     

\author{
    \IEEEauthorblockN{\hpcaauthors{}}
      \IEEEauthorblockA{
        \hpcaaffiliation{} \\
        \hpcaemail{}
      }
}

\fancypagestyle{camerareadyfirstpage}{%
  \fancyhead{}
  
  \fancyhead[C]{
    \ifdefined\aeopen
    \else
      \ifdefined\aereviewed
      \else
      \ifdefined\aereproduced
      \else
    \fi 
    \fi 
    \fi 
    \ifdefined\aeopen 
    \fi 
    \ifdefined\aereviewed
    \fi 
    \ifdefined\aereproduced
    \fi
  }
  \fancyfoot[C]{}
}
\begin{document}
\maketitle

\ifdefined\hpcacameraready 
  \thispagestyle{camerareadyfirstpage}
  \pagestyle{empty}
\else
  \thispagestyle{plain}
  \pagestyle{plain}
\fi

\newcommand{\hpcaheight}{0mm}
\ifdefined\eaopen
\renewcommand{\hpcaheight}{12mm}
\fi


\begin{abstract}

Evaluating architectural ideas on realistic workloads is increasingly challenging due to the prohibitive cost of detailed simulation and the lack of portable sampling tools. 
Existing targeted sampling techniques are often tied to specific binaries, incur significant overhead, and make rapid validation across systems infeasible. 
To address these limitations, we introduce Nugget, a flexible framework that enables portable sampling across simulators, hardware, architectural differences, and libraries. 
Nugget leverages LLVM IR to perform binary-independent interval analysis, then generates lightweight, cross-platform executable snippets, \emph{nuggets}, that can be validated natively on real hardware before use in simulation. 
This approach decouples samples from specific binaries, dramatically reduces analysis overhead, and allows researchers to iterate on sampling methodologies while efficiently validating samples across diverse systems.

\end{abstract}

\section{Introduction}
\label{sec:introduction}

Evaluating application performance on novel architectural ideas is key to computer architecture research.
However, especially for hardware which does not yet exist, evaluating application performance can be very time consuming.
For instance, when using cycle-level software simulation, there can be over 100,000$\times$ slowdown.
Even using FPGA-accelerated simulation is significantly slower than \revision{wall-clock} time~\cite{firesim}.
At the same time, modern applications are growing in their computation demands from exascale applications~\cite{exascale} to machine learning training (GPT-4 required 2,150,000 Exa- or $2.15 \times 10^{25}$ operations to complete~\cite{epoch2024trackinglargescaleaimodels}).

To facilitate the evaluation of large applications, there has been a long history of reducing applications to smaller benchmarks, kernels, or short representative intervals.
For instance, Livermore Loops provide a set of microbenchmarks that represent key computational kernels from scientific applications~\cite{livermore_loops}.
It is also common to pull out kernels from an application or to create ``proxy applications''~\cite{exascale-proxy-apps}.
While these methods can reduce very large applications to much more manageable benchmarks, they require application experts to manually identify the important parts of applications and create new programs.

Another approach to reducing the size of applications is to \emph{sample} the application to automatically find \emph{representative intervals} of different application phases.
Many techniques (i.e. SimPoint~\cite{simpoint}, BarrierPoint~\cite{barrier_point}, and LoopPoint~\cite{looppoint}) follow this approach.

In this paper, we focus on targeted sampling techniques, which use profiling to automatically select samples.
Prior targeted sampling techniques have three main drawbacks that limit their applicability to modern, realistically sized applications:

\begin{itemize}
    \item \textbf{Samples are expensive to find:} Targeted sampling techniques require running the application from beginning to end once (typically via simulation or slow emulation) to define intervals, analyze program phases, and identify representative samples.
    \item \textbf{Samples are tied to a single executable binary:} These techniques generate samples that are specific to a particular binary, so any binary change necessitates repeating the entire sampling process.
    \item \textbf{Validating the selected samples is infeasible:} Targeted sampling techniques require running the application in detailed simulation to verify that the selected samples accurately represent the program.
\end{itemize}

\revision{
To address these limitations, we present a framework called \emph{Nugget} for agile targeted sampling development.
}
Because modern applications exhibit varied characteristics, we believe there is no one-size-fits-all sampling methodology or selection algorithm~\cite{survey_phase_classification,accl_warmup_sampled_simulation,characterizing_and_comparing_prevailing_simulation_techniques,analysis_of_statistical_sampling_in_microarchitecture_simulation}.  
\revision{
Therefore, Nugget is designed to be flexible and extensible, allowing researchers to create new sample selection algorithms that can be tailored to their targeted workload and input without worrying about the current limitations of targeted sampling techniques.
}
Additionally, using Nugget for sample selection methodologies, because of its nature of cross-platform compatibility and its ability to run on real hardware, enables richer insights into the selected samples and the systems that run the samples.
The main contributions of this paper are:

\begin{itemize}
    \item \textbf{Efficient and portable interval analysis} on real hardware using LLVM, unconstrained by the hardware's microarchitecture or ISA.
    \item \textbf{An LLVM IR level unit of work}, enabling sample selection algorithms to create and locate samples for a program in a cross-platform manner, agnostic of machine level instructions, such as architectural specific instructions and optimizations.
    \item \textbf{A sample creation methodology} that allows samples to run on any platform (e.g., real hardware or simulators).
    \item \textbf{An efficient validation methodology} \revision{that validates the selected samples for the target workload and input on native hardware, integrated into the development workflow.}
\end{itemize}


\section{Motivation and Prior Work}\label{sec:motivation_and_prior_work}
Nugget is designed to address the following limitations of prior sampling techniques:

\begin{itemize}
    \item Samples are expensive to find.
    \item Samples are tied to a single executable binary.
    \item Validating the selected samples is infeasible.
\end{itemize}

In this section, we discuss each of these limitations and how prior work in sampling has or has not addressed them.

\subsection{Samples Are Expensive To Find}\label{sec:finding_samples_is_expensive}

Targeted sampling techniques require analyzing the entire workload execution, which is a time-consuming process.

For example, the SPECspeed suite in the \revision{SPEC~CPU2017} benchmark suite~\cite{spec2017} has an average of 21.8 trillion dynamic instructions per benchmark for floating point benchmarks with reference input size, which is the only input size designed to report time metrics~\cite{spec2017-workload-characterization}.
Functional simulators, such as the gem5 atomic model, are often limited to a simulation rate of approximately 1 Million Instructions Per Second (MIPS)~\cite{sustainable_gem5_simulations}.
Therefore, it takes approximately 250 days of \revision{wall-clock} time to analyze an average SPECspeed benchmark with reference input size.
More recent workloads (e.g., the forthcoming SPEC~CPUv8~\cite{spec2017v8}, HPC workloads, and machine learning workloads) are significantly larger, making functional simulation infeasible.

For multi-threaded benchmarks, the analysis overhead is even worse.
We must analyze the program once for every different thread configuration because the number of threads affects the program's execution.
Also, if the workload undergoes any changes (e.g., new compiler, different optimizations, etc.), we must redo the analysis.

Live sampling tools like Pac-Sim~\cite{pacsim} parallelize the phase analysis and detailed simulation to reduce the overall phase analysis penalty.
However, live sampling still relies on the simulator to analyze the program's execution so this technique is constrained by the simulation speed.
If a benchmark takes the simulator 250 days to analyze, as mentioned in the above example, then parallelizing the analysis and detailed simulation will still take at least 250 days.

Tools like DynamoRIO~\cite{dynamorio} and the Intel PinPoints toolkit~\cite{pinpoints} can analyze the program's execution with dynamic binary instrumentation or binary translation.
The instrumentation and profiling routines introduce overhead to the program's execution, but this overhead is significantly smaller than that of simulation (e.g., $4\text{-}40\times$ for QEMU with some single-threaded workloads~\cite{dynamicBinaryInstrumentationForBBVGeneration}).
However, these tools often have restrictions on the architectures they can analyze or the host machines.
Developing architectures that need to be analyzed with a wider range of benchmarks, such as RISC-V, often cannot use these tools because they are not supported.

\begin{tcolorbox}
\textbf{Goal 1:}
Analyze the program's execution quickly and without any restrictions on the architecture or host machines.
\end{tcolorbox}

\subsection{Samples Are Tied To A Single Executable Binary}\label{sec:samples_dont_generalize}

Most prior work relies on binary-specific information, such as the program counter, to define intervals, identify program phases, and select samples.
Any binary change, such as recompiling with different ISA extensions or compilers, alters the static instructions and the dynamic instruction stream, invalidating previously selected samples.
As discussed in Section~\ref{sec:finding_samples_is_expensive}, the high overhead of this analysis makes repeating it for every binary revision time-consuming.


Cross-binary simulation points~\cite{cross_binary_simpoints} represent a notable attempt to decouple samples from a specific binary.
However, they are limited to single-threaded programs and require significant manual effort, including mapping their control flow graphs and recalculating sample weights.

\begin{tcolorbox}
\textbf{Goal 2:}
Automatically generate samples that are independent of the binary.
\end{tcolorbox}

\subsection{Validating The Selected Samples Is Infeasible}\label{sec:validating_the_selected_samples_is_infeasible}

Sampling methodologies aim to predict overall program performance by measuring only a subset of representative samples and extrapolating the full-program behavior using statistical models. 
Most methodologies focus on predicting metrics that \revision{summarize} the entire hardware performance, such as instructions per cycle (IPC) or total runtime.

To evaluate the accuracy of a set of samples selected by a sampling methodology, the predicted performance metrics are compared against ground truth measurements. 
First, the target performance metric is measured from a full execution of the program on the experimental system, serving as the ground truth. 
Next, the same metric is measured for the selected samples, and statistical processing is applied to estimate the full-program value. 
Finally, the predicted value is compared to the ground truth to quantify the accuracy of the selected samples.

\revision{
The effectiveness of a sample selection method depends on benchmark characteristics. }
\shepherd{Differences in control flow phase behavior, working set size and locality, and input sensitivity can cause a method that succeeds on one workload and input pair to misrepresent another~\cite{characterizing_and_comparing_prevailing_simulation_techniques,simpoint,hot-region-spec2017,xalancbmk,survey_phase_classification,eeckhout-microarchitecture-independent-signatures,MAV}.}
\revision{
For example, empirical studies (e.g., MAV~\cite{MAV}) report outliers where SimPoint fails to capture distinct memory access behavior while SimPoint reports average 3\% IPC error with SPEC~CPU2000~\cite{spec2000} programs~\cite{wavelet-based}. 
Hence, representative intervals should be validated.
However, the only way to validate them is to run the entire workload.
}

\shepherd{Prior sampling techniques are typically evaluated on small benchmarks or truncated runs because their validation depends on simulation~\cite{wavelet-based,taskpoint,simpoint,looppoint,smarts,fsa,cross_binary_simpoints,pacsim,barrier_point,simpoint-plus,parrot}.}
\revision{
A missing piece is an agile path from methodology development to real hardware level validation. 
The intent of the sample set is to approximate the behavior of the real system, so validation by simulation is only meaningful if the simulator faithfully models the target. 
If it does not, the validation does not transfer.
This also leads to difficulty in comparing different methodologies as results are tied to different simulators, forcing redundant engineering effort before we can make apples-to-apples comparisons.
}
For example, LoopPoint is validated with \revision{SPEC~CPU2017} benchmarks using train-size input with Sniper~\cite{sniper}.  
FSA~\cite{fsa} is validated with the first 30~billion instructions of the \revision{SPEC~CPU2006~\cite{spec2006}} benchmarks with gem5~\cite{11_gem5,20_gem5}.
These techniques can only be validated in a limited way because they require a full detailed simulation from beginning to end of the workload to validate.
Detailed simulation can be 2--10$\times$ slower than functional simulation, which implies that validation of full-sized workloads could take nearly a decade! 

\revision{
Lack of an agile workflow to validate quickly has led to sampling techniques being rarely validated with large benchmarks.
}
\shepherd{Evidence suggests that validation results from smaller benchmarks or partial executions do not generalize to larger benchmarks and realistic workloads~\cite{characterizing_and_comparing_prevailing_simulation_techniques,hot-region-spec2017,simpoint,spec2017-workload-characterization,wavelet-based}.}
\revision{
Moreover, the program behavior of a workload with different input can be different~\cite{xalancbmk}.
Thus, per benchmark and input pair validation is needed.
}

\begin{tcolorbox}
\textbf{Goal 3:}
Quickly validating the selected samples for the targeted benchmarks and input size.
\end{tcolorbox}

\section{The Nugget Framework}\label{sec:the_nugget_framework}

Nugget is designed to address the limitations of the prior sampling techniques, specifically focusing on 1) the high cost of finding samples, 2) the coupling of the samples to a specific executable binary, and 3) the infeasibility of validating the selected samples.
Nugget leverages LLVM passes to statically insert annotations to analyze the program's execution on real hardware without the need for simulation, emulation, or binary translation.
By using LLVM intermediate representation (IR) to define program progress and intervals, the samples become independent of the binary and can be reused across different executable versions of the same program, since LLVM IR abstracts away details of the final binary.
Lastly, Nugget provides a method to create portable samples to be executed on any platform, such as real hardware or simulators, so that the samples can be validated in native time with real hardware before running them on a simulator.

\begin{figure}[!t]
    \centering
    \includegraphics[width=\columnwidth]{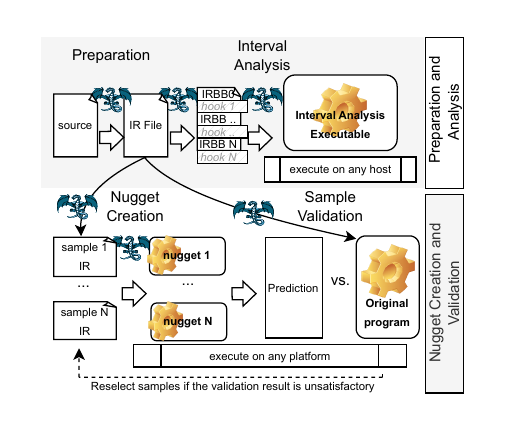}
    \caption{Nugget pipeline. The process is divided into two main stages:
    (1) preparation and analysis, and
    (2) nugget selection and validation.
    The preparation and analysis stage, including base IR file creation and interval analysis execution, only needs to be performed once per program.
    Nugget creation and validation can be repeated to refine and select the most representative set of samples for the workload.}
    \label{fig:nugget_pipeline}
\end{figure}

In this section, we present an overview of Nugget as a pipeline (Figure~\ref{fig:nugget_pipeline}), describing its workflow from preparation through interval analysis, nugget creation, and finally, sample selection verification.
It is not the focus of this paper to introduce a new sample selection algorithm, but rather to provide a framework that can be used to implement and validate diverse sample selection methodologies that can be tailored to specific applications and workloads.

\revision{We begin by defining program progress and intervals, which ensures that samples remain independent of the binary, as described in Section~\ref{sec:unit_of_work}.}
Next, we describe the preparation required by Nugget to guarantee the LLVM IR of the program remains consistent across different binaries in Section~\ref{sec:initial_setup}.
Then, we introduce our novel interval analysis methodology, which leverages LLVM passes, enabling comprehensive program analysis directly on real hardware without imposing hardware-specific constraints in Section~\ref{sec:interval_analysis}.
Following that, we present the nugget creation process, which allows for the generation of portable samples that can be executed on any platform, including real hardware and simulators, in Section~\ref{sec:nugget_creation}.
Finally, we describe the sample selection validation process, which enables efficient validation of the selected samples for specific workloads in native time, ensuring that the samples accurately represent the program's execution in Section~\ref{sec:verification}.

\FloatBarrier

\subsection{Unit of Work}\label{sec:unit_of_work}

A unit of work defines the metric by which progress toward completing a program's execution is measured. 
It must remain consistent across identical runs of a program with the same input. 
While machine instructions could serve this purpose for a single binary, Nugget instead counts executed LLVM IR instructions as its unit of work. 
This ensures a uniform measurement of program progress across all binaries of the same program with the same input.

A unit of work directly determines the notion of intervals. 
For example, if the unit of work is the number of executed instructions, then each interval corresponds to a fixed number of executed instructions during runtime. 
A sample is a specific interval of a program's execution; thus, the unit of work also defines what constitutes a sample.

The choice of unit of work influences not only how intervals are defined, but also the sample selection algorithm and the statistical models used in performance prediction. 
Therefore, the unit of work must be defined consistently across all binaries to ensure that samples remain independent of any specific binary.
For example, if the unit of work is the number of executed machine-level instructions, then the sample will be tied to a specific binary because any binary changes can affect the number of machine-level instructions executed, even if the control flow remains the same~\cite{cross_binary_simpoints}.

LLVM~\cite{llvm} is an open-source compiler infrastructure that transforms source code into a platform-independent intermediate representation (IR) before generating machine-specific binaries.
LLVM IR instructions abstract away the underlying machine-level instructions, so as long as the control flow remains the same at the LLVM IR level, the definition of all intervals of program execution remains consistent across different binaries.
Since LLVM IR instructions are an abstraction of machine-level instructions, the number of executed LLVM IR instructions reflects the progress of program execution just as the number of executed machine-level instructions does.

\shepherd{
Therefore, Nugget's unit of work is suitable for applications that have stable control flow across different platforms.
Applications such as gradient descent, where floating-point precision might affect the control flow, are not suitable for Nugget or for any sampling methodologies that rely on the application having stable control flows across microarchitectures~\cite{wavelet-based,simpoint,looppoint,cross_binary_simpoints,pacsim,barrier_point,simpoint-plus,parrot}.}

\subsection{Preparation}\label{sec:initial_setup}

In the preparation stage, we generate the optimized LLVM IR representation of the program, which is used consistently throughout all stages of the pipeline (Figure~\ref{fig:nugget_pipeline}). 
This is because Nugget relies on a stable LLVM IR representation of the program for its unit of work, as introduced in Section~\ref{sec:unit_of_work}.

Optimization is important for the program's performance.
LLVM IR level optimizations, such as loop unrolling, can significantly affect control flow.
To maintain consistency, Nugget applies these optimizations during the generation of the base LLVM IR representation in the preparation stage.
In contrast, most backend optimizations, such as replacing scalar instructions with AVX vector instructions, do not alter the structure of the LLVM IR.
While certain backend passes (e.g., late-stage loop unrolling) may modify LLVM IR layout under specific conditions, such passes are not enabled by default in LLVM.

\shepherd{While build-time modes such as LTO and PGO can alter LLVM IR, they are disabled by default.}
\revision{For our experiments, the LLVM IR structure remained stable across optimization levels, CPU-feature flags, and LLVM versions.
}
Therefore, we can apply the backend optimization in the final binary compilation safely.

Only code included in this base LLVM IR representation of the program will be subject to analysis.
For instance, any linked libraries must also be processed through Nugget during this preparation stage; otherwise, their contents will be excluded from the analysis.
This selective inclusion also allows us to ignore irrelevant or non-contributory operations, such as spin loops in multithreaded libraries~\cite{looppoint}, that do not meaningfully impact program progress.

\subsection{Interval Analysis}\label{sec:interval_analysis}

Nugget uses LLVM passes to automatically annotate programs at the LLVM IR level, enabling interval analysis to be performed directly on real hardware.
Interval analysis is necessary to discover and profile all intervals in a program's execution as defined by the chosen unit of work.
However, profiling an entire program can be time-consuming if done via simulation or emulation (Section~\ref{sec:finding_samples_is_expensive}).

While tools that rely on dynamic instrumentation and binary translation exist to perform such analyses, they often impose restrictions on the ISA and the host platform.
By embedding interval analysis directly into the program's execution, Nugget achieves near-native speed without introducing hardware-specific constraints: any machine capable of running the program can execute the interval analysis.

Because our unit of work is defined at the LLVM IR level and is independent of the final binary, all discovered intervals remain consistent regardless of the host machine's ISA or micro-architecture.
For example, we can analyze a program on an x86 system and then use the resulting intervals to drive sampling on an \revision{Arm} system.

In this section, we describe (1) how Nugget uses LLVM passes to generate an interval analysis executable, and (2) how this executable is used to discover and profile intervals.

\subsubsection{Interval Analysis LLVM Pass}\label{sec:interval_analysis_llvm_pass}


The interval analysis LLVM pass, a modular compiler transformation that operates directly on LLVM IR, identifies all basic blocks (IRBBs) in the program's IR.
For each IRBB, it records the number of LLVM IR instructions and inserts a hook at the end of the block.
Each hook is uniquely associated with its IRBB and encodes both the block's ID and instruction count, allowing us to retrieve LLVM IR-level information during binary execution.

The hook is a lightweight function that counts occurrences and invokes other functions when specified conditions are met.
This design allows us to flexibly adjust what data to collect by simply changing the functions the hook calls.
After compilation with this LLVM interval analysis pass, the resulting binary executes the hook every time an IRBB is completed, enabling runtime data collection at the LLVM IRBB level.
This hook-based approach also lets us automatically map LLVM IR basic blocks to their corresponding machine-level basic blocks.
We discuss the overheads introduced by the hooks in Section~\ref{sec:overhead_of_hooks}.

\subsubsection{Interval Analysis Executable}\label{sec:interval_analysis_executable}

The interval analysis process has two main goals: (1) to partition the program's execution into a sequence of intervals, and (2) to generate a characteristic signature (e.g., a basic block vector) for each interval.
In Nugget, an interval is defined by a fixed number of executed LLVM IR instructions (Section~\ref{sec:unit_of_work}).
Thus, discovering intervals involves recording the current execution point whenever the LLVM IR instruction count is reached and marking the current execution point as the end of a newly identified interval.

Once intervals are identified, each is characterized by a signature.
Whereas the interval's definition specifies its location in the program's execution, the signature captures its behavioral characteristics.
For example, SimPoint defines intervals by instruction count but represents each interval's signature as a basic block vector that tracks how often each basic block is executed.

A signature is closely tied to the chosen sample selection algorithm.
Since this paper does not propose a new selection algorithm, we do not introduce a new signature.
However, we demonstrate how our hook-based approach can flexibly collect different types of signatures by constructing both LLVM IRBB vectors and count stamp vectors.
The LLVM IRBB vector tracks the frequency of entering each IRBB, while the count stamp vector records the global IR instruction counter value when each IRBB is last entered.
These will later be used to create markers (Section~\ref{sec:nugget_creation}).

\paragraph*{Discovering Intervals}

Discovering intervals is managed by the hooks.
Each hook knows the ID and instruction count of its IRBB.
Whenever a hook executes, it increments a global IR instruction counter by the IRBB's instruction count and checks whether this counter has reached the defined interval size.
If so, a new interval is marked by assigning it a sequential ID; otherwise, execution continues as normal.

All discovered intervals are binary independent because their definitions rely on the LLVM IR-level.
Thus, regardless of the ISA or microarchitecture of the machine running the interval analysis executable, the discovered intervals remain consistent for the final compiled binary.

\paragraph*{Generate A Characteristic Signature}

Our hook-based design makes it easy to specify what data to collect during execution by simply modifying the functions called by each hook.
Because each hook has access to the LLVM IR basic block (IRBB) information, we can build an \emph{IRBB vector} by incrementing a global vector based on the IRBB ID whenever a block is entered.
This vector effectively records how many times each IRBB is executed within an interval, capturing the frequency profile of the program's control flow.
It is recorded and reset whenever a new interval is detected.
Since the LLVM IR control flow remains consistent across all final binaries (Section~\ref{sec:initial_setup}), the IRBB vector is inherently binary independent.

In contrast, the \emph{count-stamp vector} records the global IR instruction count each time an IRBB is entered, marking when during the program's execution each IRBB was last encountered.
This provides a timestamped trace of the control flow relative to instruction execution.
The count-stamp vector is also recorded and reset at the end of each interval.

In Section~\ref{sec:markers}, we show how the IRBB vector is used to create execution markers and how the count-stamp vector helps reduce the hook overhead.
This flexible design allows researchers to explore additional signatures for sampling methodologies with minimal development effort.

\subsection{Nugget Creation}\label{sec:nugget_creation}

After selecting a set of samples with a sampling algorithm, Nugget generates \textit{nuggets} based on these samples.
We define a \textit{nugget} as an executable that marks the selected sample with a start point and an end point, enabling execution on any platform capable of running the original program, including both real hardware and simulators.
Thus, every \textit{nugget} requires a cross-platform start marker and end marker to indicate the boundaries of the sample execution.
In this section, we describe how Nugget creates these markers and mitigates their overhead.

\subsubsection{Markers}\label{sec:markers}

In Nugget, a \textit{marker} is a mechanism for identifying a specific point in program execution, defined according to our unit of work.
Because progress is measured by the number of executed LLVM IR instructions (Section~\ref{sec:unit_of_work}), our markers are directly tied to the IR instruction count.

\shepherd{We designate the last executed LLVM IR basic block (IRBB) that completes the target instruction count as a \emph{marker}.
This marker consists of two components: the IRBB itself and the number of times it must execute. 
The end marker for an interval is determined using the IRBB vector and count-stamp vector gathered during interval analysis (Section~\ref{sec:interval_analysis_executable}).
Specifically, we use the count-stamp vector (Section~\ref{sec:interval_analysis_executable}) to identify the last executed IRBB of the interval, which becomes the basic block for the end marker, and the IRBB vector to determine how many times that block must execute to reach the interval's endpoint.
Because the start of one interval corresponds to the end of the previous interval, we reuse the end marker of the preceding interval as the start marker for the next.}

\shepherd{
Nugget inserts hooks at these marker IRBBs during the nugget creation LLVM pass.
These hooks track IRBB executions and trigger specific actions, such as resetting statistics in a simulator or starting a timer on real hardware.
}

When using \textit{nuggets} in simulation, we can also mark the start of a warmup interval.
The warmup interval does not belong to the selected sample but is a period of execution used to warm up the microarchitectural state in simulation models, assuming functional simulation is used to fast-forward execution before the sample starts~\cite{ComputerArchitecturePerformanceEvaluationMethods}.
\revision{
Any contiguous and antecedent intervals of the sample interval can be used as the warmup interval, so one can choose how many intervals to use for warmup.
} 

\subsubsection{Overhead of Hooks}\label{sec:overhead_of_hooks}

Hooks inserted at marker IRBBs introduce overhead due to frequent counting and condition checks, which can affect measurements if executed often.
This overhead appears when running \textit{nuggets} on real hardware, as the hooks add extra instructions.

For single-threaded programs, the overhead is typically small because hooks perform only simple operations, such as counting and checking thresholds.
Moreover, hooks stop all counting and checks once the threshold is reached, and modern branch predictors generally handle these conditional checks efficiently.

For multi-threaded programs, the overhead can be more significant because counting and threshold checks require synchronization (e.g., atomic operations).
Although hooks still cease checking after meeting the threshold, the overhead from the end marker's hook is unavoidable.

To mitigate this, we propose two strategies:
\begin{itemize}
    \item Identifying a lower-overhead marker.
    \item Using the program counter to track the marker instead of hook instrumentation when running in simulation.
\end{itemize}

\paragraph*{Lower Overhead Marker}

To reduce overhead on real hardware, we can trade some precision by selecting an IRBB near the end of the interval that executes less frequently than the final IRBB.
Because hook overhead is tied to execution frequency, this choice reduces cost.
We do this by defining a search distance in IR instructions from the interval's end, identifying candidate IRBBs within this distance using the count-stamp vector, and selecting the least frequently executed one using the IRBB vector.
This effectively marks a point slightly before the interval's true end, reducing overhead at the cost of some precision.

\paragraph*{Program Counter Tracking in Simulation}

While hook instrumentation is necessary on real hardware, when running \textit{nuggets} in a simulator we can leverage simulator capabilities to eliminate hooks entirely.
The LLVM nugget creation pass labels the marker IRBBs in the assembly, and we use the program counter addresses of these labels as inputs to the simulator to track markers.
This completely removes marker overhead in simulation, ensuring measurements remain unaffected.

\subsection{Sample Validating}\label{sec:verification}

Nugget addresses the limitations of using simulation to validate selected samples (Section~\ref{sec:validating_the_selected_samples_is_infeasible}) by generating \textit{nuggets} that can run directly on both real hardware and simulation.
Therefore, we can validate the selected samples in real hardware, and use them for simulation.
This approach allows for fast, scalable validation of sampling accuracy without the high overhead of simulation, even for large applications.

Because nuggets are \textbf{binary-independent}, they can be validated across different systems with minimal constraints. 
Validating the same samples on multiple platforms enhances confidence in their \textit{representativeness and robustness}. 
As shown later in Section~\ref{sec:model_performance_evaluation}, we found that having \textit{consistent prediction error across platforms} is a stronger indicator of sample quality than achieving low error on a single platform.

To run a nugget on real hardware, the system must still execute the portion of the program leading up to the sample's start marker (Section~\ref{sec:markers}).
For samples that are in later stages of a large program's execution, this can introduce some validation overhead. 

In the future, we plan to reduce this overhead further by exploring techniques such as creating software checkpoints using CRIU~\cite{CRIU} \revision{or BLCR~\cite{BLCR}} to resume execution closer to the sample start point.

\section{Evaluation}\label{sec:evaluation}

In this section, we revisit the goals of Nugget and evaluate how well we achieve them.

\begin{enumerate}
    \item Analyzing program execution quickly and without restrictions on architecture or host machine.
    \item Automatically generating samples that are independent of the binary.
    \item Quickly validating the selected samples for the targeted benchmarks and input size.
\end{enumerate}

We begin by evaluating the first goal through measurements of the analysis overhead introduced by Nugget, comparing it to both uninstrumented execution and to functional simulation (Section~\ref{sec:interval_analysis_efficiency}). 
We then address the second and third goal by employing two simple sample selection methodologies, random sampling and LLVM IR basic block (IRBB) vector-based k-means sampling, and evaluating their effectiveness (Section~\ref{sec:sampling_methodology_validation}).

\begin{table}[!t]
  \centering

  \begin{minipage}{\columnwidth}
    \centering
    \scriptsize
    \setlength{\tabcolsep}{4pt}
    \renewcommand{\arraystretch}{1.2}

    \begin{tabular}{|c|c|c|c|}
      \hline
      \textbf{Model} & \textbf{Core} & \textbf{Cache Hierarchy} & \textbf{Memory} \\ \hline
      \makecell[l]{AMD Ryzen\\Threadripper 3960X}
        & \makecell[l]{24-core OOO\\@ 3.8 GHz}
        & \makecell[l]{L1d:32 KiB L1i:32 KiB\\L2:512 KiB L3:128 MiB}
        & \makecell[l]{DDR4 2133 MHz\\136.5 GiB/s} \\ \hline
      \makecell[l]{Ampere Altra}
        & \makecell[l]{160-core OOO\\@ 2.8 GHz}
        & \makecell[l]{L1d:64 KiB L1i:64 KiB\\L2:1 MiB SLC:64 MiB}
        & \makecell[l]{DDR4 3200 MHz\\409.6 GiB/s} \\ \hline
    \end{tabular}

    \caption{Real Hardware Configuration}
    \label{tab:system-configuration}
  \end{minipage}

  \vspace{0.8em}

  \begin{minipage}{\columnwidth}
    \centering
    \scriptsize

    \begin{tabular}{|p{1cm}|p{2.5cm}|p{2.5cm}|p{1.1cm}|}
      \hline
      \textbf{Suite} & \textbf{Workload} & \textbf{Parallelism} & \textbf{Lang.} \\ \hline
      \makecell[l]{SPEC \\ CPU2017}
        & \makecell[l]{Desktop and server \\workloads}
        & Single-threaded
        & \makecell[l]{C, C++, \\ Fortran} \\ \hline
      \makecell[l]{NPB}
        & HPC kernels
        & Multi-threaded (OMP)
        & Fortran, C \\ \hline
      LSMS
        & \makecell[l]{First-principles \\ materials simulation}
        & Single-threaded
        & C++ \\ \hline
    \end{tabular}

    \caption{Overview of Benchmark Suites Used in Evaluation}
    \label{tab:benchmark-suites}
    
  \end{minipage}

\end{table}
\FloatBarrier 

Table~\ref{tab:system-configuration} presents the detailed configurations of the real hardware used in our evaluation. 
For the simulation models, we employ a basic generic functional model using the ATOMIC CPU in gem5, running a guest Ubuntu~24.04 environment, to perform interval analysis on both x86 and Arm architectures.

\revision{
We evaluate Nugget using the SPEC CPU~2017 benchmark suite, the NAS Parallel Benchmarks (NPB) suite~\cite{npb}, and the LSMS benchmark~\cite{LSMS}, covering a wide range of realistic workloads as summarized in Table~\ref{tab:benchmark-suites}. 
}

\revision{We use the ``-O2'' optimization level for the compilation.}
The input chosen for each benchmark is too large to be simulated in feasible time (except for NPB class ``A'').

For SPEC~CPU2017, we use the reference input sizes, which require between 198 billion and 9{,}897 billion instructions.
For NPB, we use class ``C'' and ``D'' inputs, with class ``C'' requiring between 24 billion and 2{,}714 billion instructions and class ``D'' ranging from 1{,}217 billion to 37{,}533 billion instructions. 
For LSMS, we use the Fe input, which requires between 28{,}484 billion and 198{,}885 billion instructions (one binary per ISA).

\shepherd{We used SPEC~CPU2017 Integer suite, except for ``619.lbm\_s'' and ``644.nab\_s'', which are from the Floating-Point suite.}
\revision{
  We do not include CPU~2017 benchmarks that failed to compile with the original SPEC~CPU2017 GCC build system (6 benchmarks) or encountered Fortran linking issues (5 benchmarks).
  We did not encounter the same Fortran linking issues with any NPB Fortran benchmarks.
}

To accurately measure hardware performance, we employ \textit{cpuset} to strictly bind workloads to specific cores and memory nodes. 
We also disable CPU frequency scaling and fix the CPU at a constant frequency. 
The LLVM passes are implemented on top of LLVM 19.0.0git.

\subsection{Interval Analysis Efficiency (Goal 1)}\label{sec:interval_analysis_efficiency}

In this section, we evaluate the efficiency of Nugget's interval analysis.
We define efficiency using two criteria:
(1) \textit{low slowdown}, quantified as the ratio between the time spent on interval analysis and the time required for executing the full workload; and
(2) \textit{flexibility}, meaning the ability to perform the analysis across different host machines.

During interval analysis, we identify all valid intervals and collect two key artifacts: the LLVM IR Basic Block (IRBB) vector and the count-stamp vector (Section~\ref{sec:interval_analysis_executable}).
These outputs are later used to generate nuggets.

For all workloads, we execute the complete preparation and analysis pipeline shown in Figure~\ref{fig:nugget_pipeline}.
Additionally, we compile an unmodified baseline binary from the same LLVM IR to isolate and measure the overhead introduced solely by the interval analysis.

We begin by comparing the overhead of interval analysis when performed via functional simulation versus Nugget.
Next, we evaluate how overhead varies across a range of workload types and use different host machines to demonstrate the flexibility of Nugget.
Finally, we study how overhead scales with the number of threads in the workload, noting that synchronization requirements during interval analysis can amplify slowdown at higher thread counts.

\subsubsection{Comparison with Functional Simulation}
\label{sec:comparison_with_functional_simulation}

For this experiment, we use the NPB suite with Class ``A'' inputs and 4 threads.
We select Class ``A'' because larger input sizes, such as those used in Figure~\ref{fig:interval_analysis_overhead}, would require months to analyze using functional simulation.
We use multi-threaded workloads to demonstrate a worst-case overhead scenario, since interval analysis must track synchronization events between threads.
Additionally, multi-threaded workloads reflect current trends in modern computing, making them a relevant evaluation target.

To evaluate the efficiency of Nugget, we compare it against LoopPoint~\cite{looppoint}, a recent state-of-the-art sampling methodology.
LoopPoint is selected for comparison because it supports multi-threaded workload sampling and has integrated support within the gem5 simulator for full-system analysis~\cite{gem5looppoint}.
This is particularly relevant, as all nuggets in our framework are generated on real hardware, a full-system environment.

\begin{figure}[!t]
    \centering
    \includegraphics[width=\linewidth]{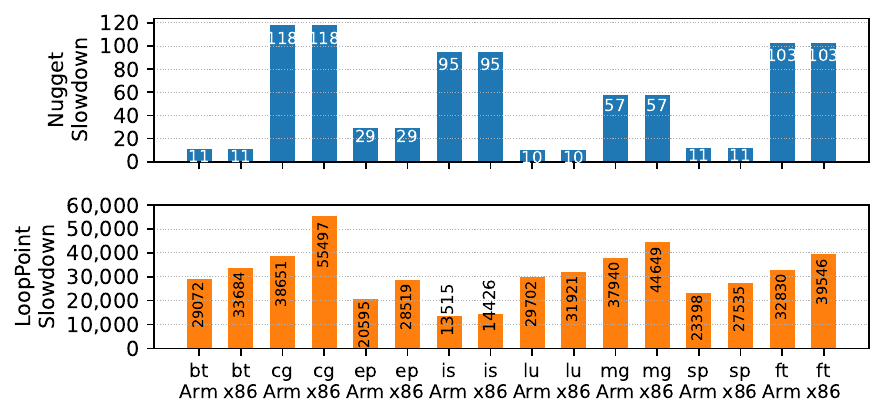}
    \caption{Interval analysis overhead via Nugget and functional simulation.}
    \label{fig:simulation_interval_analysis_overhead}
\end{figure}

As shown in Figure~\ref{fig:simulation_interval_analysis_overhead}, we observe an average $577.65\times$ reduction in interval analysis overhead when using Nugget instead of gem5's functional simulation.
The overhead on the order of tens of thousands in functional simulation highlights its impracticality for interval analysis in large applications.
On average, Nugget incurs a slowdown of $54\times$, with a maximum of $118\times$ for the ``cg'' benchmark.
By comparison, LoopPoint's average overhead is $31,\!343\times$, peaking at $55,\!497\times$ for ``cg-x86.''

Overall, multi-threaded workloads amplify interval analysis overhead because their baseline execution times are significantly \revision{shorter (due to parallelism) while} the analysis itself incurs greater cost from synchronization requirements.
The single-threaded workloads in Figure~\ref{fig:interval_analysis_overhead} exhibit much lower overheads, which we discuss further in Section~\ref{sec:mix_of_workload_types}.
The ``cg'' benchmark has the shortest execution time in this experiment (0.28 seconds), which further amplifies the measured overhead for both methods.
The ``is'' benchmark exhibits the lowest overhead under functional simulation but still incurs a $95\times$ slowdown with Nugget, primarily due to the frequent executions of interval analysis hooks, as discussed in Section~\ref{sec:hook_overhead}.
The ``mg'' and ``ft'' benchmarks show relatively high overhead for both methods because their parallel regions increase synchronization costs during interval analysis, which in turn adds further simulation overhead.
We examine the relationship between synchronization and overhead in more detail in Section~\ref{sec:scaling_with_number_of_threads}.

In addition to its lower overhead, Nugget offers a key practical advantage: it requires only a single interval analysis run to generate nuggets that can be executed across multiple architectures.
In contrast, LoopPoint must repeat interval analysis for every binary variant or target architecture, significantly increasing its total cost.

\subsubsection{Mix of Workload Types}
\label{sec:mix_of_workload_types}

\begin{figure}[!t]
    \centering
    \includegraphics[width=\linewidth]{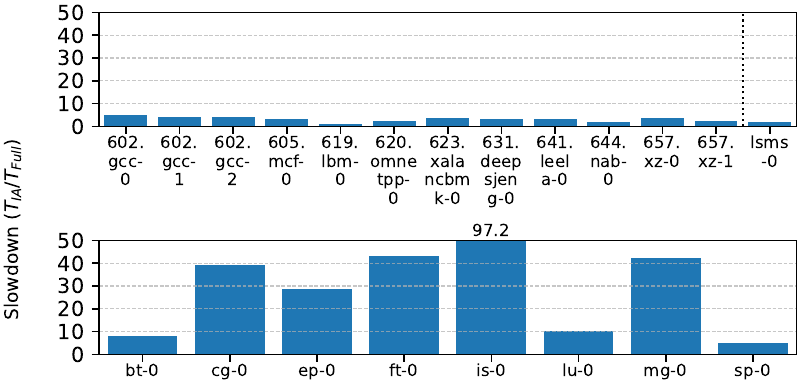}
    \caption{Interval analysis overhead via Nugget across different workloads.}
    \label{fig:interval_analysis_overhead}
\end{figure}

We also evaluate overhead across a mix of workload types.
For single-threaded workloads, we use the SPEC~CPU2017 suite with the \texttt{ref} input size and the LSMS benchmark with the \texttt{Fe} input size.
For multi-threaded workloads, we use the NPB suite with Class ``C'' in this experiment.

Figure~\ref{fig:interval_analysis_overhead} shows the interval analysis slowdown across the evaluated workloads and demonstrates that \textbf{the overhead varies significantly depending on the level of synchronization}.
The average observed overhead is \(3\times\) for SPEC~CPU2017, \(2\times\) for LSMS, and \(34\times\) for NPB.
The overhead remains low for SPEC~CPU2017 and LSMS because they are single-threaded workloads, which simplifies interval analysis by eliminating the need for synchronization.
In contrast, the high overhead for NPB arises from the synchronization required during multi-threaded interval analysis, particularly due to aggregation across threads.
As synchronization demands increase, so does the overhead, a relationship we explore further in Section~\ref{sec:scaling_with_number_of_threads}.

\textbf{To demonstrate the flexibility of our approach across host machines, we repeat the interval analysis on two distinct systems listed in Table~\ref{tab:system-configuration} and find consistent results across both platforms for single-threaded workloads, with one corner case.}
For SPEC~CPU2017, interval analysis results are identical across the two systems.
For NPB, the multi-threaded nature of the benchmarks introduces variability between runs (even on the same host), a well-known challenge in multi-threaded sampling that is beyond the scope of this work~\cite{looppoint,barrier_point}.
For LSMS, we observe minor differences in the execution frequency of some IRBBs, which we attribute to differences in floating-point precision across CPUs.
In particular, the number of iterations in certain data-driven loops in LSMS is influenced by the underlying architecture's precision.
Such differences could affect nugget representativeness if a marker IRBB is sensitive to these variations.
Nevertheless, as shown in Section~\ref{sec:sample_validation}, nuggets generated from interval analysis on one system still achieve low prediction error when validated on the other system for LSMS.

\subsubsection{Scaling with Number of Threads}
\label{sec:scaling_with_number_of_threads}

\begin{figure}[!t]
    \centering
    \includegraphics[width=\linewidth]{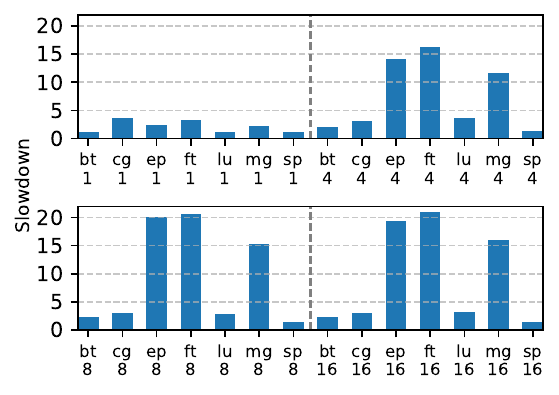}
    \caption{Interval analysis overhead via Nugget across different numbers of threads.}
    \label{fig:class_d_overhead_analysis}
\end{figure}

Finally, we study how interval analysis overhead scales with the number of threads.
For this experiment, we use the NPB suite with Class ``D'' inputs and vary the thread count across \{1, 4, 8, 16\}.
To ensure fair comparisons, all overheads are normalized to the execution time of the original workload running with a single thread.

As shown in Figure~\ref{fig:class_d_overhead_analysis}, most workloads exhibit increasing overhead with higher thread counts, consistent with our expectation that more threads lead to greater synchronization overhead.
However, this growth is not uniform across all workloads because not all of them require the same level of synchronization during interval analysis, as not all hooks are executed within parallel regions.
For ``bt'', ``cg'', ``lu'', and ``sp'', the consistently low slowdown as the thread count increases suggests that hooks are not frequently executed within parallel regions for these benchmarks.
In contrast, ``mg'' and ``ft'' show a significant increase in overhead with more threads, indicating that these benchmarks require more synchronization during interval analysis---likely due to the complexity of their parallel regions.

Even at higher thread counts (up to $16\times$), Nugget maintains overheads that are orders of magnitude lower than those of functional simulation-based interval analysis (Figure~\ref{fig:simulation_interval_analysis_overhead}).

\subsubsection{Summary}\label{sec:interval_analysis_overhead_summary}

Our experiments show that Nugget achieves orders-of-magnitude lower interval analysis overhead compared to functional simulation, with an average reduction of over $577.65\times$ relative to LoopPoint for multi-threaded workloads.
It also produces consistent analysis results across different host machines, demonstrating its portability.

By enabling fast, binary-independent interval analysis, even for large workloads, Nugget successfully meets Goal~1.

In the following sections, we build upon this foundation to demonstrate how Nugget supports a variety of sampling methodologies (Goal~2) and enables rapid validation of selected samples on real hardware (Goal~3), further establishing its effectiveness for sampling methodology research.

\subsection{Applying Sampling Methodologies to the Nugget Framework and Validating the Selected Samples (Goals 2 \& 3)}
\label{sec:sampling_methodology_validation}

\begin{figure*}[t]
    \centering
    \includegraphics[width=\textwidth]{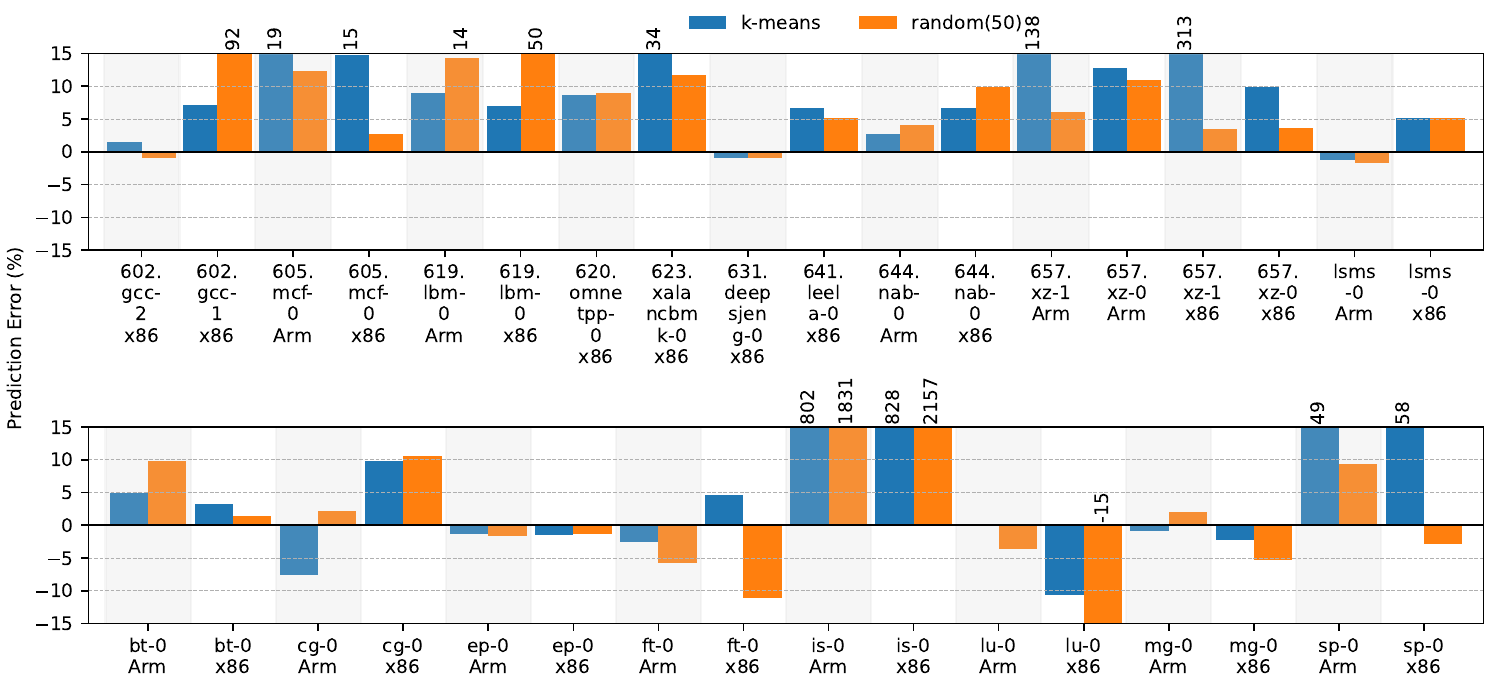}
      \caption{
        Prediction error by machine and sampling method.
        \textbf{All results shown are obtained from executing the nuggets directly on real hardware}, without using simulation.
    }
    \label{fig:prediction_error_by_machine_method}
\end{figure*}

The primary objective of Nugget is to support diverse sample selection methodologies while providing efficient interval analysis, binary-independent samples, and scalable validation capabilities.
In this section, we leverage these features to execute the selected samples, \emph{nuggets}, directly on real hardware, enabling us to assess whether different sampling strategies can accurately predict full workload performance \emph{without relying on simulation}.

Because this paper does not propose new sampling algorithms, we focus on two straightforward approaches: random sampling and LLVM IRBB vector-based k-means clustering.
We show that Nugget seamlessly implements these methodologies and facilitates validating the selected samples on \emph{real hardware}.
We also show that different workloads may benefit from different sampling strategies; Nugget enables discovering and applying methods tailored to specific workload characteristics.
Moreover, we highlight the importance of validating sampling methodologies across multiple machines, especially when the target system is not yet available.
Finally, we discuss the overhead introduced by the hooks used to mark sample intervals within nuggets.

\subsubsection{Sampling Methodologies}
\label{sec:sampling_methodology}

For the random sampling approach, we randomly select a set of intervals from the program execution to estimate its average or cumulative properties.
This selection method is based on statistical sampling methods~\cite{smarts}.

For the clustering approach, we use the k-means algorithm to cluster the IRBB vectors, select ``k'' samples to represent the program phases of the workload, and use the cluster weights to estimate the average or cumulative properties of the program execution.
This approach is based on basic block vector (BBV) phase-identification techniques~\cite{survey_phase_classification} (e.g. SimPoint) and on the fact that the IRBB vector is an abstraction of the actual BBV.

We refer to the random approach as ``Random'' and the clustering approach as ``K-means''.
We use 50 samples as the target number for both methodologies, which is statistically large enough to approximate the population distribution~\cite{hogg2020psi}.
We apply the Silhouette score with the K-means algorithm to determine the optimal number of clusters for each workload, enforcing the constraint $\#\text{clusters} \le 50$ (with $k = 50$ as the upper bound).
We assign equal weights to all intervals in the ``Random'' approach, whereas the ``K-means'' approach weighs each interval according to the size of its cluster.

\subsubsection{Sample Validation}
\label{sec:sample_validation}

We use the analysis information collected in Section~\ref{sec:mix_of_workload_types} to select samples with the ``Random'' and ``K-means'' approaches.
We then generate nuggets for the selected samples and validate them as described in the Sample Creation and Validation step in Figure~\ref{fig:nugget_pipeline}.
The nuggets are used to predict the runtime of each workload, with the runtime of the full execution serving as the baseline for calculating prediction error.
We choose runtime as the metric because it is a common target of sampling methodologies~\cite{looppoint}, and IPC has been shown to be unsuitable for multi-threaded workloads~\cite{ipc_considered_harmful}.

Figure~\ref{fig:prediction_error_by_machine_method} presents the prediction errors of nuggets generated by the Random and K-means approaches for each workload on the real-hardware machines listed in Table~\ref{tab:system-configuration}. 
As shown, many workloads exhibit significant errors, often exceeding the accuracy reported by prior methodologies such as SimPoint (3\% for single-threaded workloads) and LoopPoint (2.9\% for 8-threaded workloads).

Several factors contribute to these elevated errors.
(1) The sampling methodologies are naively designed. For example, the Random approach relies heavily on luck and the specific composition of the interval pool.
\shepherd{
(2) Previous work rarely validates samples on real hardware~\cite{wavelet-based,taskpoint,simpoint,looppoint,smarts,fsa,cross_binary_simpoints,pacsim,barrier_point,simpoint-plus,parrot}.
Instead, they mostly rely on user-mode-only simulation, which is typically less noisy due to more controllable thread interleaving compared to actual hardware~\cite{variability_in_architectural_simulations_of_multithreaded_workloads,11_gem5,survey-simulator-techniques,determineistic_shared_memory_multiprocessing}. 
In Section~\ref{sec:model_performance_evaluation}, we further analyze the sources of prediction error using smaller workloads in simulation.
(3) Prior work rarely validates large workloads as we used in our experiments (e.g., reference input sizes for CPU2017). 
Larger and more realistic applications exhibit increasingly complex behaviors~\cite{characterizing_and_comparing_prevailing_simulation_techniques,hot-region-spec2017, simpoint,spec2017-workload-characterization,wavelet-based,looppoint}, which demand more representative sample sets, as discussed in Section~\ref{sec:validating_the_selected_samples_is_infeasible}. 
}
In such cases, large program phases can amplify prediction error.
(4) Hardware-based nuggets incur additional overhead for the begin and end interval markers, particularly in multi-threaded contexts, which further inflates prediction error. We provide a detailed discussion of this overhead in Section~\ref{sec:hook_overhead}.

As the figure shows, the sampling approach that yields the lowest prediction error varies by workload.
For example, for ``sp,'' the ``Random'' approach with 50 samples achieves the lowest prediction error across all methods, and this holds on both machines, suggesting that this sample set is architecturally and microarchitecturally independent.
In contrast, for ``lbm,'' the ``K-means'' approach consistently yields the lowest prediction error.

We focus on demonstrating Nugget's ability to validate samples on distinct machines in this section.
In Section~\ref{sec:model_performance_evaluation}, we will further analyze how to utilize the hardware validation data with a case study.
We also find that maintaining consistency across machines is a stronger indicator of sample quality than minimizing prediction error on any single platform.
This insight further underscores the importance of efficient, cross-platform validation in the design of robust sampling methodologies.

\subsubsection{Hook Overhead}
\label{sec:hook_overhead}

\begin{figure}[!t]
    \centering
    \includegraphics[width=\linewidth]{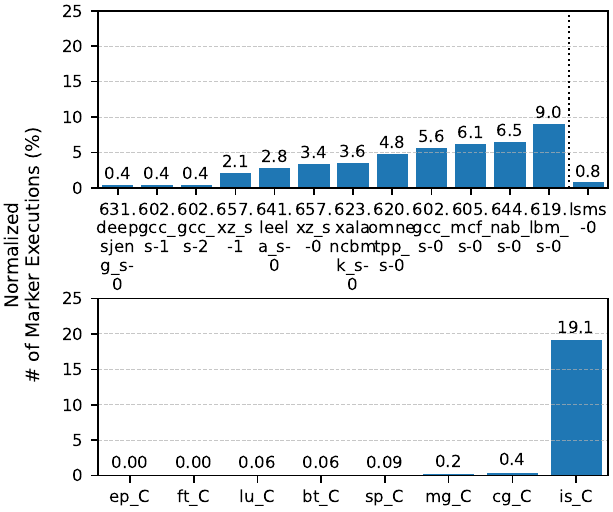}
    \caption{\shepherd{The number of marker hook executions for the randomly selected nuggets shown in Figure~\ref{fig:prediction_error_by_machine_method}. The number of hook executions is normalized to the total number of IRBB executions during all nugget executions for each workload and input pair.}}
    \label{fig:marker_executions_combined}
\end{figure}

Nugget uses hook instructions to mark sample intervals, which can introduce runtime overhead during workload execution.
\shepherd{
In Figure~\ref{fig:marker_executions_combined}, we present the number of times the marker hooks are executed for the randomly selected nuggets shown in Figure~\ref{fig:prediction_error_by_machine_method}.
The number of hook executions is normalized to the total number of IRBB executions during all nugget executions for each workload and input pair.
As shown, ``is'' exhibits a significantly higher number of hook executions compared to other workloads, indicating that the prediction error observed for ``is'' in Figure~\ref{fig:prediction_error_by_machine_method} is influenced by the excessive overhead from hook executions.
}
Because ``is'' frequently executes only a small number of IRBBs within a parallel region, it triggers an excessive number of hook invocations.
``is'' was also one of the workloads with the largest overheads for analysis as shown in Figure~\ref{fig:simulation_interval_analysis_overhead}.
This leads to substantial overhead and results in inaccurate prediction errors.
\shepherd{Hook overhead can only slow execution, so it can cause false negatives (rejecting good samples) but is unlikely to cause false-positive conclusions.}

\shepherd{
For other workloads, the number of hook executions does not show a clear correlation with the prediction error observed in Figure~\ref{fig:prediction_error_by_machine_method}.
Cleanly separating the impact of hook overhead from other sources of prediction error is still challenging.
The number of hook executions and the prediction errors observed suggest that the cutoff line for a good marker should be executing lower than 10\% for single-threaded workloads and lower than 1\% for multithreaded workloads, in terms of the total IRBBs executed in the nugget, to avoid significant overhead impacting the validation results.
Multithreaded workloads are more sensitive to hook overhead due to the synchronization required for counting and threshold checks, as discussed in Section~\ref{sec:overhead_of_hooks}.
}

This behavior highlights a limitation of the current hook-based interval marking approach.
Future work could explore alternative mechanisms to reduce this overhead or to decouple it from the validation results.
However, addressing this issue is nontrivial, especially for multithreaded workloads: overhead in parallel regions can differ significantly from that in serial regions, and parallelism may sometimes mask or amplify hook-related costs, complicating the isolation of their impact.

It is important to note that this overhead is unique to real hardware execution.
In simulation, as discussed in Section~\ref{sec:overhead_of_hooks}, sample intervals can be marked directly by the simulator without introducing runtime overhead.

\section{Case Studies}\label{sec:case_study}

In this section, we present two case studies.
First, we separate the overhead of nugget markers from the accuracy of the sampling selection process by running the nuggets both on hardware and in simulation.
Then, we illustrate how nuggets can serve as lightweight microbenchmarks to assess simulation model accuracy, eliminating the need to develop dedicated microbenchmark suites.

\subsection{Nuggets in Simulation and on Hardware}\label{sec:model_performance_evaluation}


The goal of computer architecture analysis is often to compare the performance of two systems.
Therefore, to investigate the usefulness of nuggets, we run both the entire workload and the nuggets and compare the speedup between different systems.
We compare four different systems \emph{in simulation} to eliminate the potential inaccuracies from the interval selection marker in hardware.
We also compare four different hardware systems.
With these eight systems, we can make two conclusions.
(1) We can use hardware to provide increased confidence in the selected intervals, but we cannot prove or bound the error due to the overheads discussed in Section~\ref{sec:case_study_simulation_results}.
(2) We find that Random and basic-block vector approaches to interval selection are not microarchitectural independent.
In fact, we find that microarchitectural differences affect the accuracy of the selection methodology more than the ISA.

In this study we report the \emph{error in predicted speedup}, not the error in absolute runtime.
We found that in cases with modest absolute error the predicted speedup (or relative performance) is often much more accurate.


\subsubsection{Experimental Setup}

\begin{table}[!t]
  \centering
  \scriptsize  
  \setlength{\tabcolsep}{3pt}
  \renewcommand{\arraystretch}{1.2}

  \begin{tabular}{|c|c|c|c|c|}
    \hline
    \textbf{Model}
      & \textbf{ISA}
      & \textbf{Core}
      & \textbf{Cache Hierarchy}
      & \textbf{Memory} \\ \hline

    Config 1
      & x86
      & \makecell[l]{4-core Large OOO\\@ 4 GHz}
      & \makecell[l]{L1d: 64 KiB L1i: 64 KiB\\L2: 1 MiB}
      & \makecell[l]{DDR4 2400 MHz\\38.4 GB/s} \\ \hline

    Config 2
      & Arm
      & \makecell[l]{8-core Small OOO\\N1 @ 3 GHz}
      & \makecell[l]{L1d: 64 KiB L1i: 64 KiB\\L2: 1 MiB SLC: 32 MiB}
      & \makecell[l]{DDR4 2400 MHz\\76.8 GiB/s} \\ \hline

    Config 3
      & x86
      & \makecell[l]{4-core Base OOO\\@ 4 GHz}
      & \makecell[l]{L1d: 64 KiB L1i: 64 KiB\\L2: 1 MiB}
      & \makecell[l]{DDR4 2400 MHz\\38.4 GB/s} \\ \hline

    Config 4
      & Arm
      & \makecell[l]{8-core Base OOO\\@ 4 GHz}
      & \makecell[l]{L1d: 64 KiB L1i: 64 KiB\\L2: 1 MiB}
      & \makecell[l]{DDR4 2400 MHz\\76.8 GiB/s} \\ \hline

  \end{tabular}

  \caption{gem5 Simulation Model Configuration. \revision{The large, small, and base out-of-order cores are internal models that target different performance characteristics.}}
  \label{tab:system-configuration-simulation}
\end{table}

We use the selected nuggets to conduct experiments on both gem5, using full-system simulation, and real hardware.
In addition to the systems listed in Table~\ref{tab:system-configuration}, we include two additional real hardware platforms: an Intel i9-9900X system and an Intel i7-9700 system.
The simulation models used are listed in Table~\ref{tab:system-configuration-simulation} and run a guest Ubuntu~24.04 kernel, matching the software environment of the real hardware.

Sample sets are selected using the ``Random'' and ``K-means'' methods described in Section~\ref{sec:sampling_methodology}.
We set $k = 20$ in this experiment, as the use of NPB Class A with a single-threaded configuration results in a relatively small pool of intervals available for sampling.
Single-threaded workloads also help eliminate confounding factors such as thread interleaving, which could otherwise affect our measurement of prediction error.


In simulation, we use gem5 to track sample interval markers directly, thereby avoiding the runtime overhead introduced by hook-based marking mechanisms.
To simulate nuggets efficiently, we first use KVM to fast-forward execution to the warmup marker and then create a gem5 checkpoint~\cite{gem5-checkpoints}.
This allows us to resume nugget execution from that point using different microarchitectural configurations, subject to certain constraints (e.g., the number of cores must remain fixed).
For each nugget, we restore from the checkpoint and begin measurement only after reaching the start marker, allowing the microarchitectural state to warm up naturally.

\subsubsection{Results}\label{sec:case_study_simulation_results}

\paragraph*{Hardware-Based Validation Offers Scalability and Early Signal}

Validating samples on real hardware is significantly faster and more scalable than using simulation.
Several simulation runs (e.g., ``bt'', ``lu'', and ``sp''), even with small inputs such as Class A, remain incomplete after two weeks, making simulation impractical for extensive validation.
By contrast, hardware validation enables us to quickly assess the variability in speedup prediction across platforms.


\begin{figure}[!t]
  \centering
  \includegraphics[width=\linewidth]{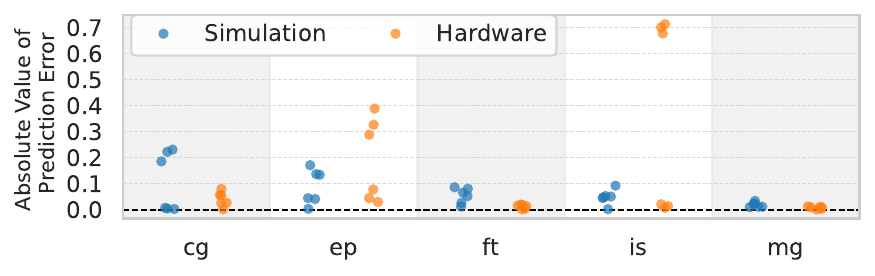}
  \caption{Distribution of absolute speedup prediction errors using K-means samples across simulation (blue) and hardware (orange) comparisons for selected benchmarks. ``0.1'' means 10\% difference in predicted speedup.}
  \label{fig:abs-speedup-prediction-error-var}
\end{figure}

Figure~\ref{fig:abs-speedup-prediction-error-var} shows the distribution of absolute speedup prediction errors for selected benchmarks.
Each point is absolute value of the error of one pair of configurations' (in simulation) or pair of hardware systems' speedup compared to the true speedup of those two systems determined by running the entire workload.
The hardware results allow us to quickly identify benchmarks with high variability in prediction error, such as ``cg'', ``ep'', and ``is'', which exhibit wide error ranges across different platforms.
For these workloads with large variations or large errors on hardware, we find the intervals are \emph{not representative of the true speedup}.
However, we cannot use the error on hardware to bound the error in simulation (e.g., the error for ft in simulation is larger than on hardware).
The main reason we cannot use the hardware error to bound the simulation error is that the error is strongly affected by the underlying microarchitecture.
In the next section, we analyze the sources of prediction error in more detail using simulation data and controlled case studies.

\paragraph*{ISA vs. Microarchitecture: Case Study}

To isolate the sources of prediction error, we compare three types of platform pairs:
(1) systems with different ISAs but small \revision{microarchitecture} differences,
(2) systems with different significantly different microarchitectural configurations, but the same ISA, and
(3) systems that differ in both ISA and microarchitecture.

\begin{figure}[!t]
  \centering

  \includegraphics[width=\linewidth]{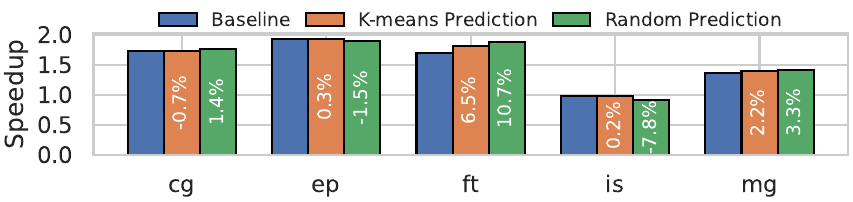}
  \caption{Speedup: Random vs K-means samples when ISA varies and microarchitecture is fixed, Config~3 vs Config~4.}
  \label{fig:case_study_3}
\end{figure}

Figure~\ref{fig:case_study_3} shows the speedup prediction error when the main difference between platforms is the ISA (Config~3 vs Config~4).
In this experiment, the ISA has a noticeable impact on the binaries: for instance, the Arm binaries have vectorization enabled, whereas the x86 binary does not.
The average absolute speedup prediction error is 5\% for Random-selected nuggets and 2.0\% for K-means-selected nuggets.
Among all benchmarks, ``ft'' exhibits the highest K-means error (6.5\%), likely due to an unrepresentative sample set.
A linear regression of nugget runtimes across the two ISAs explains 94.5\% of the variance, indicating that the selected samples capture similar execution phases across architectures.
The K-means speedup error for ``ft'' is consistent with its runtime prediction errors on Config~3 (4.9\%) and Config~4 (--1.5\%), suggesting the overestimation stems from sample imbalance.
This implies that certain performance-critical phases were not captured by the selected nuggets.
The Random-selected samples show a similar pattern for both the ``ft'' and ``is''.

\begin{figure}[!t]
  \centering
  \includegraphics[width=\linewidth]{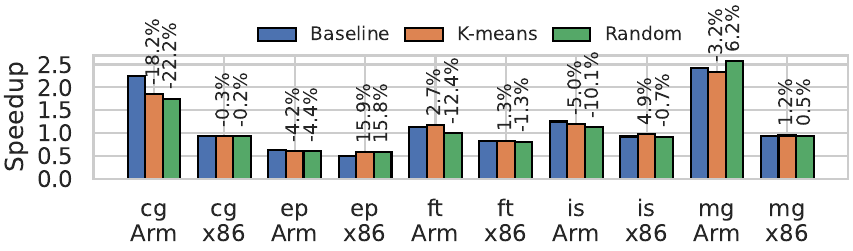}
  \caption{Speedup between Random and K-means samples when microarchitecture varies and ISA is fixed: Config~4 over Config~2 for Arm, and Config~3 over Config~1 for x86.}
  \label{fig:case_study_4}
\end{figure}

Figure~\ref{fig:case_study_4} shows the speedup prediction error when the microarchitecture varies: Config~4 over Config~2 for Arm and Config~3 over Config~1 for x86.
The average absolute prediction error increases to 8.1\% for Random-selected nuggets and 5.9\% for K-means-selected nuggets.
The comparison between Config~4 and Config~2 for Arm results in higher error in predicted speedup using the selected nuggets than between Config~3 and Config~1 for x86, reflecting greater microarchitectural differences in the Arm systems despite running the same benchmarks and sample sets.

\begin{figure}[!t]
  \centering
  \includegraphics[width=\linewidth]{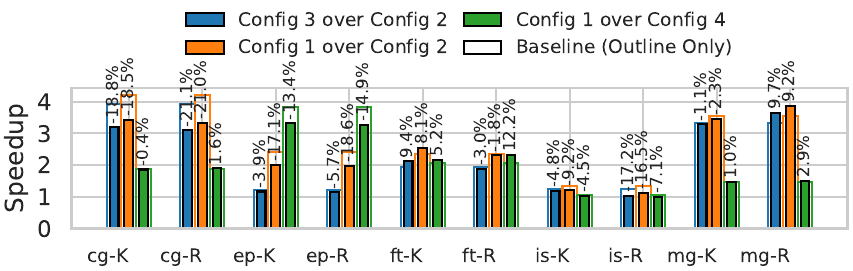}
  \caption{Speedup comparison between Random and K-means samples when both ISA and microarchitecture vary across simulation models.
  ``K'' indicates K-means samples, and ``R'' indicates Random samples.}
  \label{fig:case_study_5}
\end{figure}

Figure~\ref{fig:case_study_5} presents the speedup prediction error when both ISA and microarchitecture vary across simulation models.
The average absolute error rises to 10.8\% for Random-selected nuggets and 7.9\% for K-means-selected nuggets.
The lowest errors occur in comparisons between Config~1 and Config~4 (7.7\% for Random, 5\% for K-means), where microarchitectural differences are relatively minor.
In contrast, larger errors are observed in comparisons involving Config~2, which diverges significantly from other configurations in both ISA and microarchitecture.
These findings highlight the sensitivity of prediction accuracy to microarchitectural differences more than to ISA variation.

\textbf{In summary}, while this paper does not propose a new method for selecting representative intervals, we show the importance of comparing different interval selection techniques across a wide variety of architectures and microarchitectures.
Nugget enables future work to have apples-to-apples comparisons across architectures and fast comparisons on hardware opening this new line of research.


\subsection{Model Accuracy Evaluation}\label{sec:model_accuracy_evaluation}

In this section, we present how we used nuggets to identify sources of errors in gem5's models.
Identifying sources of errors in the simulation model is key to establishing baseline performance for evaluating architectural ideas.

Microbenchmarks are a common tool for fine-tuning architectural simulators.
To build a useful model of the Ampere Altra processor, we developed a collection of models including CPU, cache, and memory models using publicly available information on the hardware~\cite{ampere-altra}.
We used a set of microbenchmarks and fine-tuned the configuration of the models to lower the error in simulated cycles~\cite{vertical-microbench}.
While we were able to reduce cycle count error to below 5\% for many of the microbenchmarks, the models struggled to accurately simulate real applications such as NAS Parallel Benchmarks---with errors exceeding 600\%~\cite{npb}.

Using microbenchmarks has two main limitations.
(1) In many cases the compiler optimizations can remove essential parts of the code, resulting in false positives.
(2) Even when preserved after compiler optimizations, these microbenchmarks often exercise a narrow slice of the design space, missing the complex interactions between microarchitectural components present in real programs.

While it is possible to develop more sophisticated microbenchmarks, this approach is labor intensive and does not scale well as architectures develop.
Nuggets are an organic alternative to microbenchmarks;
they capture the diverse instruction mix and control flow found in the real workloads and they are small enough to provide a focused window for manual inspection by the users.


To improve our model's accuracy, we collected hardware performance counters for 234 nuggets and compared the hardware counters to gem5 statistics for each nugget to identify the nuggets with the largest error.
We found the data cache statistics explained most of the runtime error and for more than 97\% of the nuggets, the error in the number of accesses to the data cache could be explained by differences in the number of memory instructions.
We isolated the problem by inspecting execution traces from a small set of the nuggets. 
We observed that the nugget with the largest error was dominated by \revision{Arm's} paired memory instructions (e.g., ldp, stp), while fewer were present in samples with small error.

\begin{figure}[!t]
  \centering
  \includegraphics[width=\linewidth]{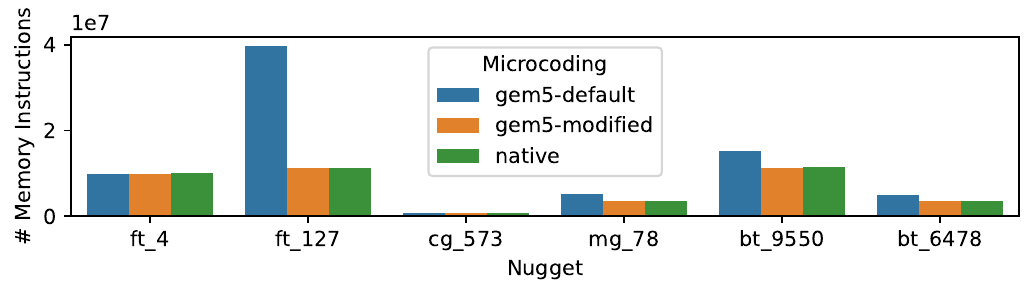}
  \caption{Comparison of the number of memory instructions for two different microcodings in gem5 vs real hardware.}
  \label{fig:microcode_comp}
\end{figure}

Figure~\ref{fig:microcode_comp} shows a comparison of the number of memory instructions in our selected nuggets for three different microcodings.
The orange bars show the number of memory instructions with the default microcoding of gem5;
the blue bars show the number of memory instructions if the paired memory instructions were microcoded as one microop;
while the green bars show the number of memory instructions measured on native hardware.
The blue bars closely track the green bars, suggesting that the microcoding of paired memory instructions in gem5 is different from the native hardware.


\section{Conclusion}\label{sec:conclusion}
In this paper we presented Nugget: a framework for building program snippets
that can be used for evaluating architectural ideas using full-system simulators.
Nugget enables automatic instrumentation of a workload's binary for analysis, and for measurement.
Binaries used for analysis can be run on real hardware much faster than prior work while
the binaries used for measurement can be run on real hardware and simulators alike.
Lastly, Nugget allows users to leverage real hardware to evaluate how well the selected samples represent the original workload.

Nugget is positioned to allow researchers to simulate workloads larger than the traditional sizes, enabling researchers to gain insights that were not possible before.

\section*{Acknowledgements}
We thank Alen Sabu, Christopher Batten, Derrick Quinn, Matthew D. Sinclair, Oscar Hernandez, and the Davis Computer Architecture Research Group for feedback used to improve this work. 
This work was supported by NSF grant 2311888, Department of Energy, Oak Ridge National Laboratory, and Los Alamos National Laboratory. 

\clearpage

\bibliographystyle{IEEEtranS}
\bibliography{refs}

\clearpage

\end{document}